\documentclass[preprint]{aastex}
\usepackage{epsfig}

\newcommand {\be} {\begin{equation}}
\newcommand {\ee} {\end{equation}}

\def\te{T_{\rm e}}

\def\scatt{d\sigma_{ij}/d\Omega}
\def\sig{\sigma_{\rm T}}
\def\b{\frac{B}{B_{\rm cr}}}
\def\bi{\frac{B_{\rm cr}}{B}}

\begin{document}

\title{The Effect of Vacuum Polarization and Proton Cyclotron
Resonances on Photon Propagation in Strongly Magnetized Plasmas}

 \author{Feryal \"Ozel\altaffilmark{1}} 
\affil{Institute for Advanced Study \\ Einstein Dr., Princeton, NJ
 08540; fozel@ias.edu }
\altaffiltext{1}{also Physics Department, Harvard University}

\begin{abstract}

We consider the effects of vacuum polarization and proton cyclotron
resonances on the propagation of radiation through a strongly
magnetized plasma. We analyze the conditions under which the photons
evolve adiabatically through the resonant density and find that the
adibaticity condition is satisfied for most photon energies of
interest, allowing for a normal-mode treatment of the photon
propagation. We then construct radiative equilibrium atmosphere models
of strongly magnetized neutron stars that includes these effects,
employing a new numerical method that resolves accurately the sharp
changes of the absorption and mode-coupling cross sections at the
resonant densities. We show that the resulting spectra are modified by
both resonances and are harder at all field strengths than a blackbody
at the effective temperature. We also show that the narrow absorption
features introduced by the proton cyclotron resonance have small
equivalent widths. We discuss the implications of our results for
properties of thermal emission from the surfaces of young neutron
stars.

\end{abstract}

\section{Introduction}

Vacuum polarization is a quantum electrodynamical phenomenon that
occurs in strong magnetic fields ($B \gtrsim 10^{13}$~G) and affects
the interactions between the photons and the electrons. In the
presence of a plasma with a density gradient, vacuum polarization
gives rise to a resonance when the normal modes of photon propagation
change from being mostly circularly polarized at high electron
densities to being mostly linearly polarized at low densities. This
occurs because, at low plasma densities, virtual pairs of the vacuum
dominate the interactions and the photon polarization eigenstates do
not correspond to the propagation eigenstates. Thus, at a critical
density that depends on photon energy, the conversion of photons
between the two polarization modes is highly enhanced, accompanied by
a change in the opacities of the normal modes (Adler 1971; Tsai \&
Erber 1975; M\'esz\'aros \& Ventura 1979; Kaminker, Pavlov, \&
Shibanov 1982; see M\'esz\'aros 1992 for a review). Another phenomenon
that affects the propagation of photons in a magnetized plasma is the
proton cyclotron resonance that arises from the interaction of the
protons in strong magnetic fields with photons.  When the field
strength is of the order of $B \approx 10^{14}-10^{15}$~G, the proton
cyclotron energy falls in the soft X-ray band and affects the spectral
properties of isolated cooling neutron stars.

Because of the sharp transition in the opacities of the normal modes
of propagation through the resonances, it has been difficult to
include the effects of vacuum polarization in calculations of photon
transport in magnetized plasmas. This phenomenon has been treated only
recently in the context of interactions of high energy photons with a
strongly magnetized plasma by Bezchastnov et al.\ (1996) and Bulik \&
Miller (1997) and included in a radiative equilibrium model atmosphere
of a neutron star by \"Ozel (2001). It has been shown that the
enhanced absorption and mode conversion give rise to broad-band
absorption features, which may affect the spectra of high-energy
bursts of soft gamma-ray repeaters (Bulik \& Miller 1997) and may be
responsible for the hard spectral tails observed in the quiescent
X-ray emission of radio-quiet neutron stars, such as the anomalous
X-ray pulsars (\"Ozel 2001). Proton cyclotron resonance has also been
addressed recently in the context of surface emission from strongly
magnetized neutron stars (Zane et al.\ 2001; Ho \& Lai 2001). However,
a treatment that takes all these effects into account has not yet been
developed (see the Appendix for the conceptual and numerical problems
with Ho \& Lai 2002). 

All these radiative transfer calculations involve solving two coupled
radiative transfer equations for the two normal modes of propagation.
Treating the vacuum polarization resonance in this formalism has some
limitations. In particular, the large Faraday depolarization limit
assumed in the derivation of the two coupled transfer equation may not
hold and an exact treatment may require the solution of four coupled
equations for the four Stokes parameters that describe the amplitudes
and phases of the electromagnetic waves (see Section 2).  However, if
the propagation modes evolve adiabatically, the normal mode treatment
is valid at densities infinitesimally away from the resonant density
and has allowed the derivation of the transfer coefficients in such
conditions (M\'esz\'aros \& Ventura 1979; Pavlov \& Shibanov 1979;
Kaminker, Pavlov \& Shibanov 1982 and references therein). All studies
to date involve this treatment, which assumes the adiabatic evolution
of the propagation modes and thus includes the enhanced conversion of
photons between the two polarization modes near the vacuum
polarization resonance. The condition of adibaticity at the resonant
density, however, needs to be verified.

In this paper, we first discuss in full generality the propagation of
photons in a magnetized plasma in the presence of resonances and
determine the conditions under which a normal mode treatment that
assumes adiabatic evolution can be employed. We then explicitly show
the terms in the radiative transfer equations that describe the
enhanced coupling of the photon propagation modes in the presence of
vacuum polarization resonance and present a new numerical treatment of
the resonances. We also take into account the additional effects of
proton cyclotron resonance and present spectra from radiative
equilibrium calculations of strongly magnetized neutron star
atmospheres when all these effects are taken into account. In an
appendix, we address various technical aspects of photon propagation
in the presence of a resonance and clarify some incorrect claims that
have recently been made in the literature.

\section{Adiabatic Evolution of Normal Modes in the Presence of 
Vacuum Polarization}

In this section, we discuss in full generality the propagation of
photons through a magnetized plasma when vacuum polarization effects
are considered. Our treatment follows closely the discussion in
M\'esz\'aros (1992, section 6.1d) and Gnedin \& Pavlov (1974). We 
show the derivation of the two coupled equations that describe the
transport of radiation in the normal modes, emphasizing the
assumptions made in this approach as well as the physical phenomena
captured in the resulting equations.

We start by defining the correlation matrix of the components 
of the electric field of the electromagnetic wave in terms of the four
Stokes parameters $I, Q, U$, and $V$:
\be
\rho_{\alpha\beta} = \frac{1}{2} \left(\begin{array}{cc} I+Q & U+iV \\ 
U-iV & I-Q \end{array} \right)\,
\ee
so that the the transfer equation that desribes its evolution takes
the form
\be
(\hat{k}\cdot\vec{\nabla}) \rho_{\alpha\beta} = -\frac{1}{2} \sum_\gamma
(T_{\alpha\gamma}\rho_{\gamma\beta}+\rho_{\alpha\gamma}T^+_{\gamma\beta}) 
+S_{\alpha\beta}.
\ee
In this equation, $T_{\alpha\beta}$ is the transfer matrix that
describes the transition from one polarization to the other as well
the absorption and outscattering of radiation, whereas
$S_{\alpha\beta}$ is the source matrix that describes emission and
inscattering processes. In the basis of eigenvectors of the transfer
matrix (which correspond to the normal modes of propagation), 
the transfer equation becomes
\be
\label{eq:full}
(\hat{k}\cdot\vec{\nabla}) R_{ij}=-g_{ij}R_{ij}+S_{ij}, 
\ee
where $R_{ij}$ and $S_{ij}$ are the projections of the correlation and 
source matrices onto the eigenvectors and 
\be
g_{ij} = \frac{1}{2}(\kappa_i+\kappa_j)+\frac{i\omega}{c}(n_i-n_j).
\ee
Here, $\kappa_i$ is the sum of the absorption and scattering
coefficients, $\omega$ is the photon frequency, and $n_i$ is the
refractive index of the $i$th mode. Note that the two equations for
the for the diagonal terms of the correlation matrix $R_{ii}$
correspond to the most general form of equation~(14) of Lai \& Ho
(2002).

When 
\be
\label{Eq:faraday}
\Im \int_z g_{ij} dz \gg \Re \int_z g_{ij} dz,
\ee
where $\Im$ and $\Re$ denote the imaginary and real parts of the
integral respectively, then the contribution of the non-diagonal
terms of $g_{ij}$ to the evolution of $R_{ij}$ is negligible because
the integral over this term oscillates rapidly.  In this case, the
transport of radiation can be described in terms of only two equations
for the specific intensity of the two normal modes
\be
(\hat{k}\cdot\vec{\nabla}) I_i=-\kappa_i I_{i}+S_{ii}, 
\ee
i.e., the familiar polarized transfer equations. Here, the term
$S_{ii}$, which takes the form
\be
S_{ii} = \sum_j \int d\Omega^\prime \frac{d\sigma_{ij}}{d\Omega}
(\Omega^\prime \rightarrow \Omega) I_j(\Omega^\prime)
\ee
when thermal effects are neglected, depends on the specific
intensities of both modes and contains information about their
coupling through the differential cross section
$d\sigma_{ij}/d\Omega$. We will discuss the properties of $S_{ii}$ and
specifically the mode coupling in the presence of vacuum polarization
resonance in the next section.

\begin{figure}[t]
\centerline{ \psfig{file=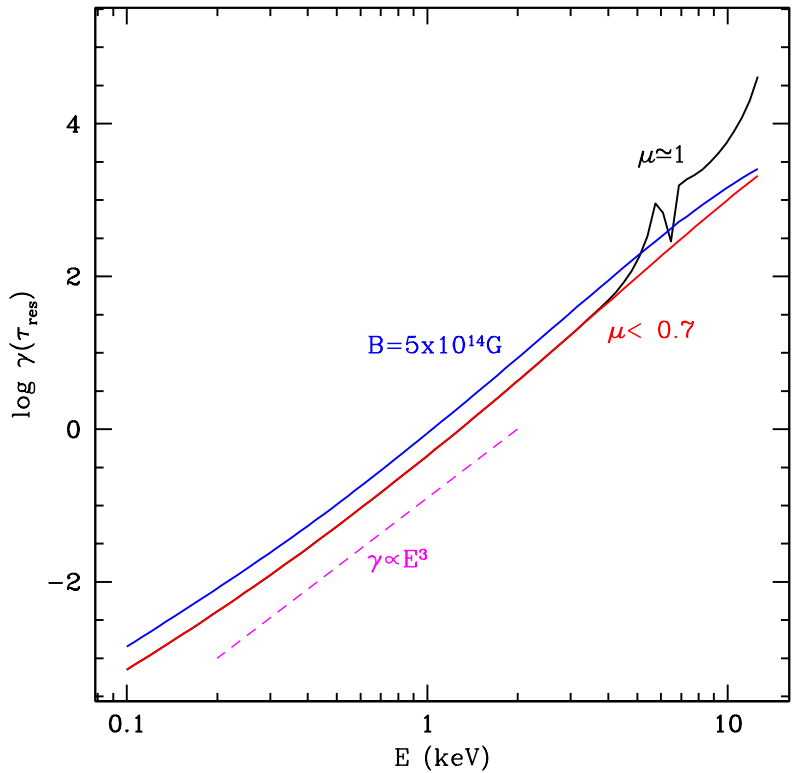,width=11truecm} } \figcaption[]{ The
quantity $\gamma$ that measures the degree of Faraday depolarization
at different photon energies, evaluated at the vacuum critical density
and the corresponding temperature in a magnetized neutron star
atmosphere with $T_{\rm eff} =0.5$~keV. The two curves labeled with
the photon direction of propagation $\mu$ correspond to $B=10^{15}$~G,
while the third curve is obtained for $\mu \lesssim 0.7$ at
$B=5\times10^{14}$~G. Since $\gamma > 1$ for $E \gtrsim 1$~keV, the
adiabaticity condition holds at these photon energies.
\label{Fig:adiab}}
\end{figure}

In a plasma with gentle density and temperature gradients, the 
condition (\ref{Eq:faraday}) becomes
\be
\label{Eq:faraday2}
\gamma \equiv \frac{\Im \int_z g_{ij} dz}{\Re \int_z g_{ij} dz}
= \frac{2 \omega (n_i-n_j)}{(\kappa_i-\kappa_j)c} \gg 1,
\ee
which is generically referred to as the limit of large Faraday
depolarization. In all the calculations of photon propagation in
magnetized plasmas to date, the normal mode treatment has been
employed at the limit of large Faraday depolarization. This requires
that the condition (\ref{Eq:faraday}) holds, i.e., that the
propagation modes evolve adiabatically through any density gradient in
the plasma, including at the vacuum resonance.  In
Figure~\ref{Fig:adiab}, we evaluate the magnitude of the quantity
$\gamma$ to verify the validity of this assumption. We use the model
atmosphere calculations discussed in \S4 for neutron stars with
magnetic fields $B=5\times 10^{14}$~G and $B=10^{15}$~G and effective
temperatures $T_{\rm eff}=0.5$~keV to calculate $\gamma$ numerically
at the resonant density and the corresponding temperature for each
photon energy. Note that the quantity $\gamma$ plotted in this figure
includes the effects of both scattering and absorption (see \S 3 for
these expressions; c.f.  equation~(16) of Lai \& Ho 2002).

As Figure~\ref{Fig:adiab} shows, the adiabaticity condition holds for
photon energies $E \gtrsim 1$~keV for all directions of photon
propagation and the magnetic field strengths of interest here, but
breaks down at smaller energies.  However, the vacuum resonance for
these low energies occurs at high optical depths in both modes (see
\"Ozel 2001), ensuring that both modes are thermalized and have the
same local radiation field density because of high number of
interactions. As a result, not including the off-diagonal terms of
$R_{ij}$ and $g_{ij}$ in the description of the radiation field at
these energies does not affect the results of the transfer
calculations.

\section{Enhanced Coupling between Normal Modes in the Presence of 
Vacuum Polarization}

Another phenomenon related to the vacuum polarization resonance in
strong magnetic fields is the drastic enhancement of the mode-coupling
terms in the source matrix $d\sigma_{ij}/d\Omega$, $i\ne j$, which
describes the redistribution of photons into different directions of
propagation and different polarization states. The derivation of the
absorption and scattering coefficients $\kappa_i$ as well as of the
matrix $d\sigma_{ij}/d\Omega$, are given in detail in, e.g.,
M\'esz\'aros \& Ventura (1979) and M\'esz\'aros (1992). The standard
procedure involves combining the sourceless Maxwell's equations into a
wave equation, casting it in the form of the transfer equation, and
identifying the various terms with the elements of the matrices
$g_{ij}$ and $S_{ij}$ (eq. [3]). Therefore, these terms take into
account {\it all\/} the mode-changing processes in a plasma, whether
or not they arise from a single scattering or, as discussed in detail
in M\'esz\'aros (1992, p.97), from non-linear effects near the vacuum
resonance (assuming adiabatic mode evolution).

Because we are specifically interested in the off-diagonal,
mode-changing terms of the photon redistribution matrix $\scatt$, we
highlight two interesting phenomena that have a significant
contribution to these terms (see also Ho \& Lai 2002). The photons
interact with both the electrons and the protons in the plasma.  When
the field strength is sufficiently high ($B \approx 10^{14-15}$~G),
the proton cyclotron energy lies in the keV range and affects the
X-ray spectra of magnetized sources. Just like the electron cyclotron
resonance, the proton cyclotron resonance gives rise to enhanced mode
coupling due to its effect on the normal modes of propagation (see
below) and to absorption features in the spectra of photons emerging
from magnetized plasmas.

The second and most interesting effect in strong magnetic fields
arises from the presence of virtual pairs which change the interaction
of the photons with the plasma and thus its index of refraction.  This
happens through the effect of the virtual pairs on the normal modes of
propagation as discussed earlier. The expressions for the absorptive
and dispersive properties of the plasma under these conditions were
derived by Adler (1971) and Tsai \& Erber (1975).

When the plasma electrons and protons as well as the vacuum
polarization effects are considered, the parameter that determines the
ellipticities of the normal modes of photon propagation take on the
form 
\be 
q=\frac{\sin^2\theta}{2\cos\theta} (1-u_p) \sqrt u
\left(1-W\frac{u-1}{u^2}\right), 
\ee 
where $u=E_b^2/E^2, u_p=E_p^2/E^2, E_b$ is the electron cyclotron
energy, $E_p$ is the proton cyclotron energy, and $E$ denotes the
photon energy. The vacuum parameter $W$ represents the correction to
the index of refraction due to vacuum polarization and has two
limiting forms; for $B < B_{\rm cr} = 4.41 \times 10^{14}$~G it is
given by
\be 
W = \frac{\alpha}{15 \pi}\left(\b\right)^2
\left(\frac{E_{\rm b}}{E_{\rm pl}}\right)^2 = 
\left( \frac{3 \times 10^{28} {\rm cm}^{-3}}{N_e}\right) \left(
\frac{B}{B_{cr}}\right)^4. 
\ee
while for $B > B_{\rm cr}$, it can be written as 
\begin{eqnarray}
 W &=& \frac{\alpha}{2\pi} \left[\frac{2}{3}(\b)-1.27+\bi \ln 
\left(2\b\right)-0.386\bi+0.70 \left(\bi\right)^2 \right. \nonumber \\
&&\quad \left. -\frac{2}{3}+\bi \ln\left(2\b\right)-0.838\bi
+\left(\bi\right)^2 \right ] 
\left(\frac{E_{\rm b}}{E_{\rm pl}}\right)^2, 
\end{eqnarray}
where we used the expansions of the electromagnetic Lagrangian 
given in Tsai \& Erber (1975, Eq. 38a-b). In these expressions, 
$\alpha$ is the fine-structure constant, $E_{\rm pl}$ is the 
plasma frequency and $N_e$ is the electron density. Given
the ellipticities of the normal modes of propagation, it is
straightforward to calculate the absorption and scattering terms 
in equation (6).

\begin{figure}[t]
\centerline{ \psfig{file=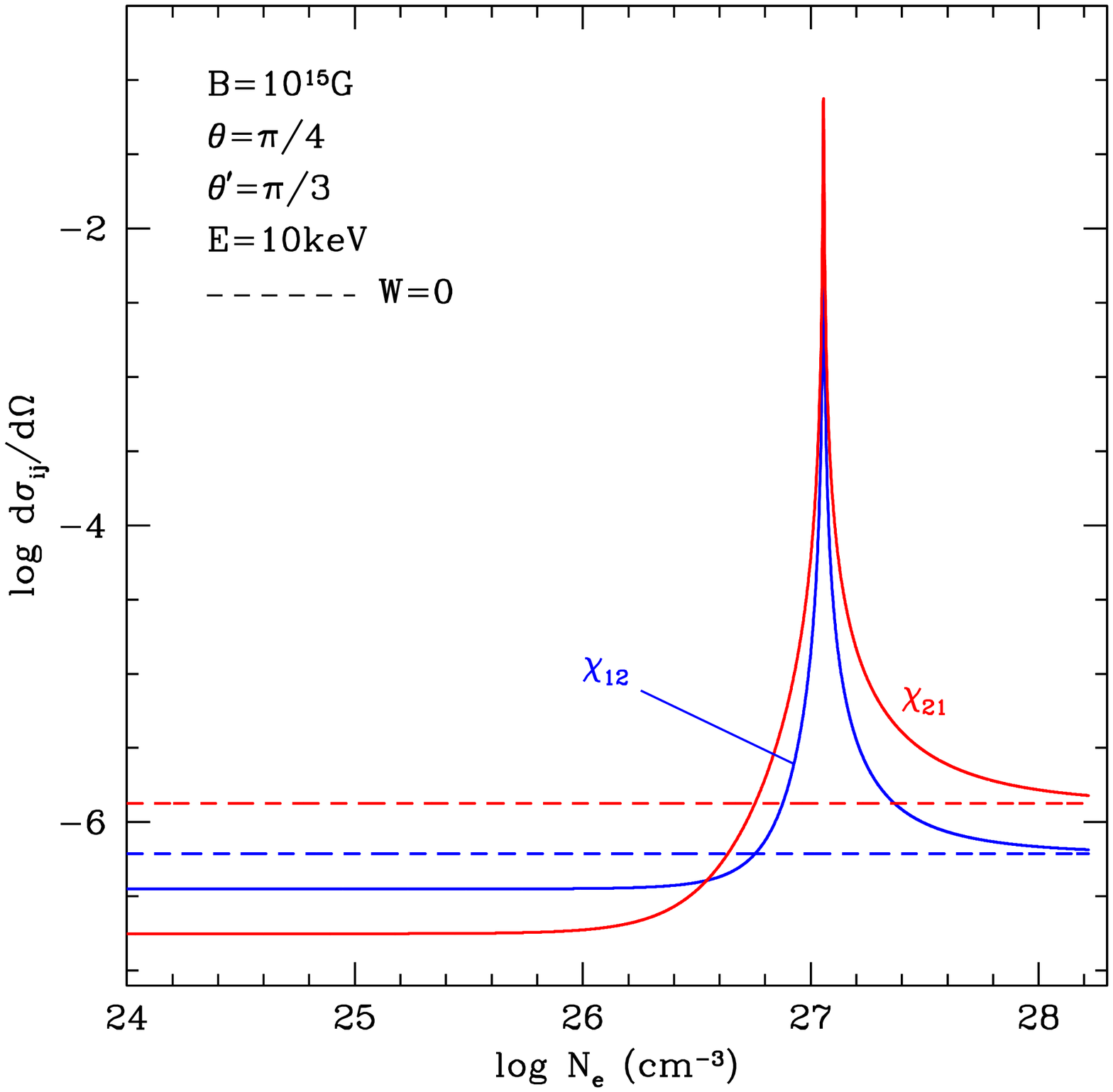,width=7.5truecm}
\psfig{file=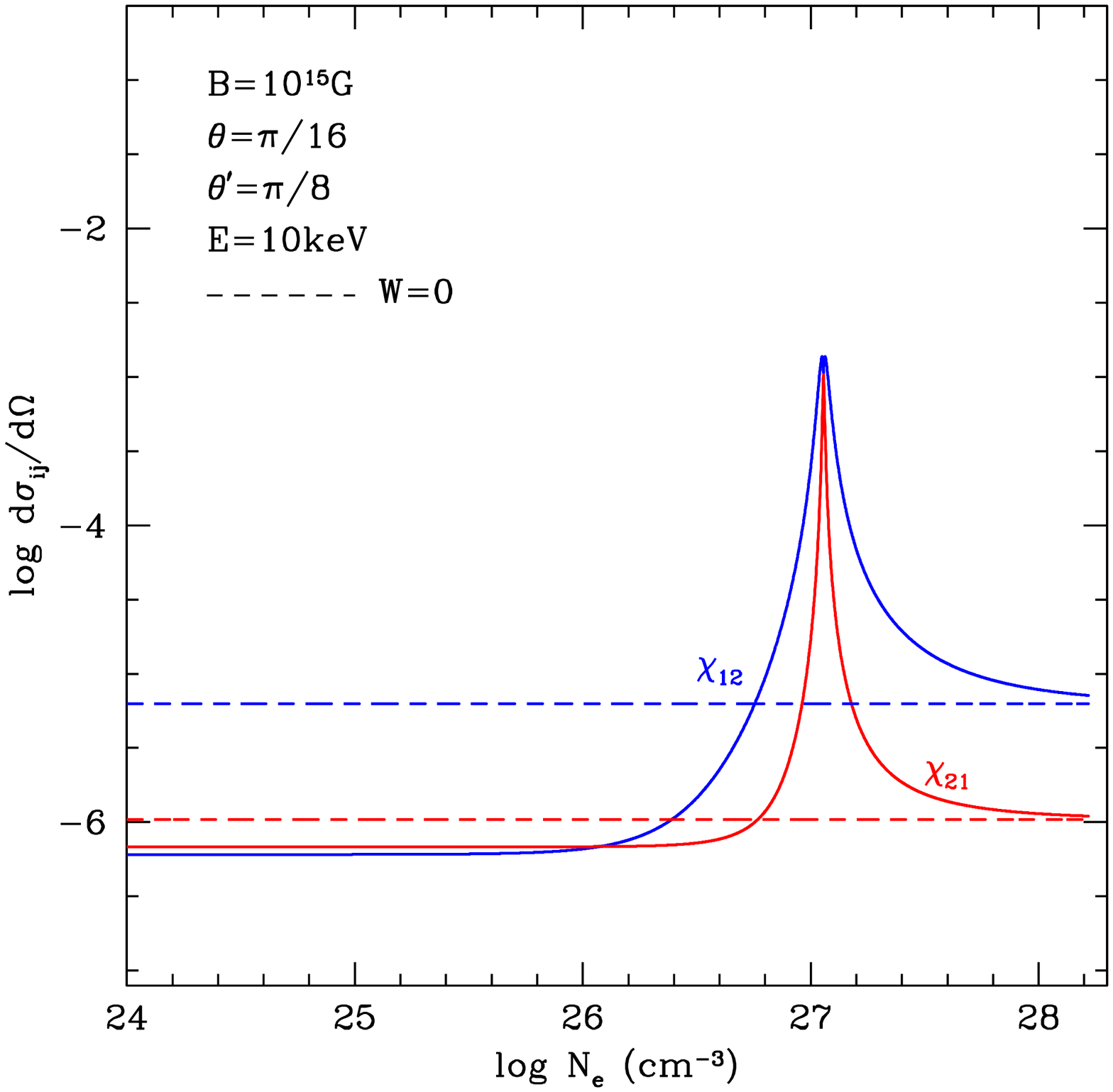,width=7.5truecm} } \centerline{
\psfig{file=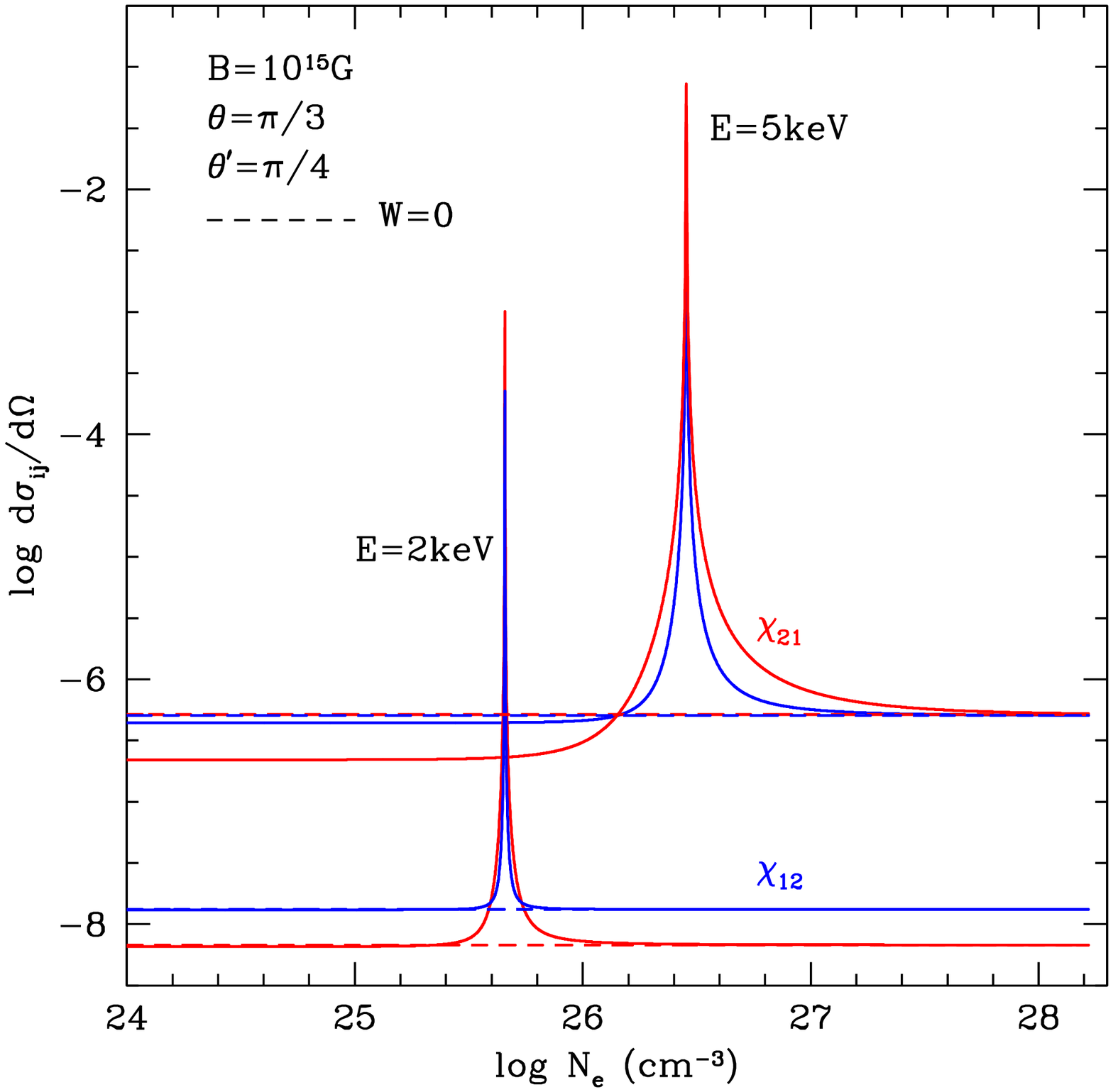,width=7.5truecm} } \figcaption[]{The
off-diagonal, mode-coupling terms of the scattering matrix $\scatt$,
in units of $N_e \sig$, as a function of particle density in the
atmosphere.  In all the panels, solid lines show the terms when vacuum
effects are included and the dashed lines when these effects are
neglected. Coupling of photons with different directions of
propagation and different energies are shown in panels (b) and (c),
respectively. \label{Fig:opac}}
\end{figure}

The quantity $q$ determines directly the polarizations of the normal
modes and thus all the elements of $\scatt$, which involve the moduli
of the cyclic projections of the unit polarization vectors onto the
coordinate axis with a given magnetic field direction. In
Figure~\ref{Fig:opac}, we plot the off-diagonal terms of $\scatt$ as a
function of electron density for different photon energies and
directions of propagation.

Figure~\ref{Fig:opac} shows a remarkable enhancement in the
off-diagonal terms of the scattering matrix near the vacuum resonance
density. These terms describe {\it true\/} propagation-mode changing
and can significantly alter the spectrum and angular distribution of
radiation propagating through a magnetized medium. This is because
they can convert a photon from a mode with a small mean-free path to
one with a large mean-free path and vice versa. As the different
panels of Figure~2 show, the conversion probability and the width of
the resonance depends on the photon energy and direction of
propagation.

These rapidly-changing off-diagonal terms are difficult to handle
numerically when modeling the transport of radiation. A number of
different approaches have been employed to date, which we discuss in
the next section.

\section{Numerical Treatments of the Resonance and Mode Coupling} 

As discussed in \S3, vacuum polarization introduces narrow and
energy-dependent resonances in the opacities and cross-mode
interaction terms. Given that the radiative equilibrium calculations
require an energy grid that extends over four orders of magnitude and
a depth grid that spans ten orders of magnitude, resolving the
resonance by using an arbitrarily large number of grid points in these
two variables is computationally prohibitive. Instead, the solution
requires new numerical methods, which can sample the resonance region
with high accuracy. The accuracy of the solutions depends on a correct
calculation of the total optical depth under the resonance.

Here we introduce a new algorithm to overcome this problem that
involves sampling the resonance region with a very large number of
points on an auxiliary grid (denoted by prime) in order to compute
accurately the total optical depth. We smooth the redistribution 
matrix elements according to
\be 
\frac{d\sigma^{\rm n}_{ij}}{d\Omega}(\tau_{\rm es}) = 
\int \frac{d\sigma_{ij}}{d\Omega}(\tau^\prime_{\rm es})
\frac{1}{\sqrt{2 \pi \sigma^2}} 
\exp\frac{-(\tau_{\rm es}-\tau^\prime_{\rm es})^2}{2\sigma^2} 
d\tau^\prime_{\rm es}\,
\ee
where $d\sigma^{\rm n}_{ij}/d\Omega$ denotes the new smoothed matrix
elements and the smoothing factor $\sigma$ in the Gaussian is chosen
based on the number of points in the main depth grid. The number of
points on the auxiliary grid can be arbitrarily large without
increasing significantly the computational time. (Note that we also
compute the absorption coefficients in this way). This method allows
the sharp features to be resolved on the discrete grid while
preserving the total optical depth across the resonance.  As a result,
it yields the most accurate solution allowed for a chosen number of
grid points, as well as smooth spectra. In the calculations presented
in \S 4, we employ a depth grid of 400 and an energy grid of 80
points.

\subsection{Comparison with Previous Methods}

We now compare the above algorithm to the numerical methods have been
developed earlier to treat the transfer of radiation through the
vacuum polarization resonance.

\noindent {\bf Monte Carlo Methods.---} Bulik \& Miller (1997) used a
Monte Carlo technique to follow the propagation of photons through a
hot, ultramagnetized plasma. Since Monte Carlo methods are not
grid-based but follow the trajectories of individual photons, they
have little difficulty in handling the sharp vacuum resonances. They
also allow an easy implementation of non-coherent (Compton)
scattering, which was important for the high temperature plasmas
considered by Bulik \& Miller (1997). This method, however, suffers
from small number statistics that produces numerical noise in the
calculated spectra.  It is also not suitable for steady-state
calculations in which the radiative equilibrium condition is required
throughout the medium, as in the case of a cooling neutron star
atmosphere.

\begin{figure}[t]
\centerline{ \psfig{file=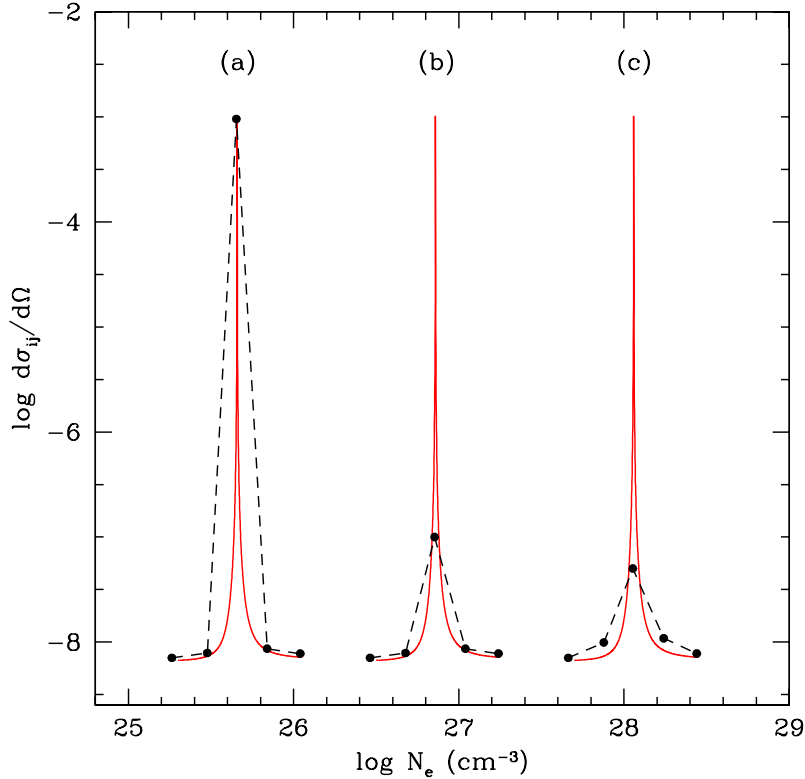,width=11truecm} } \figcaption[]{A
schematic representation of three different grid choices for resolving
a vacuum resonance feature: (a) the equal-grid method of Ho \& Lai
(2002), (b) the saturation method of \"Ozel (2001), and (c) the
Gaussian-smoothing method presented here.
\label{Fig:methods}}
\end{figure}

\noindent {\bf Grid-based Methods.---} Grid-based methods facilitate
the construction of atmosphere models in steady-state in which the
radiative equilibrium condition is imposed. Such methods are not
constrained by small number statistics and can resolve sharp features
given a carefully chosen grid. In the presence of vacuum polarization
and cyclotron resonances, the choice of grid is crucial in capturing
the effects discussed in \S3.  A typical resonance in the off-diagonal
terms of the scattering matrix is illustrated in
Figure~\ref{Fig:methods}, along with three different choices of grid
that have been recently employed.

Ho \& Lai (2002) used a scheme, the equal-grid method, in which for
every energy grid point, the vacuum resonant density corresponds to a
depth grid point (Fig.~3, case a). This choice ensures a uniform
sampling of all resonant features and thus the smoothness of the
resulting spectra. However, because the vacuum features are very
narrow, such a choice always overestimates significantly the total
optical depth across the resonance as is evident in the
figure. Indeed, even though the calculated spectra are smooth, they
have artificial ``absorption-like'' features that are reduced as the
number of grid points increases (see Fig.~9 of Ho \& Lai 2002). Note,
however, that even at the highest number of grid points used by Ho \&
Lai (2002), the artificial features are still present in the spectra
which have not approached the true solution. It is likely that this
effect is also responsible for the discrepancy at low energies between
the ``no-conversion'' and ``mode-conversion'' solutions presented in
Figures~11 and 12 of Ho \& Lai (2002).

\"Ozel (2001) approached this problem by devising a saturation scheme
which truncates the sharp resonant features (Fig.3, case b). In this
method, a large number of grid points in column depth are used to
ensure that for every energy, at least one depth grid point samples
the saturated value of the resonant feature. This method
underestimates the optical depth across the resonance but, for the
same number of grid points, the error introduced is much smaller than
the equal grid method (see Fig.~3). It also converges faster to the
true solution as the number of grid points increases. Note that in
\"Ozel (2001), a saturation value and the corresponding number of grid
points were chosen such that the solutions reached an asymptotic limit
and did not depend on the particular choices.

Note, however, that the computed optical depth under the vacuum
feature, which determines the accuracy of the solutions, in not
preserved in either method discussed above. This is contrast to the
new method presented in this paper, which is shown as case (c) in 
Figure~3.

Finally, the ``step-function'' method involves matching and exchanging
the opacities of the two polarization modes below and above the
resonance without resolving the transition region. It was first
introduced by Zane et al.\ (2001) in the case of ultramagnetized
neutron star atmospheres and is also used in Ho \& Lai (2002) in their
``complete mode conversion'' calculations. This approximate method is
highly inaccurate and provides no advantages over the equal-grid
method in resolving the sharp change of the absorption coefficients
near the resonance as it equally misestimates the total optical depth
across the resonance.

\section{The Effect of Vacuum Polarization and Proton Cyclotron Resonances
on Radiation Spectra}

\begin{figure}[t]
\centerline{ \psfig{file=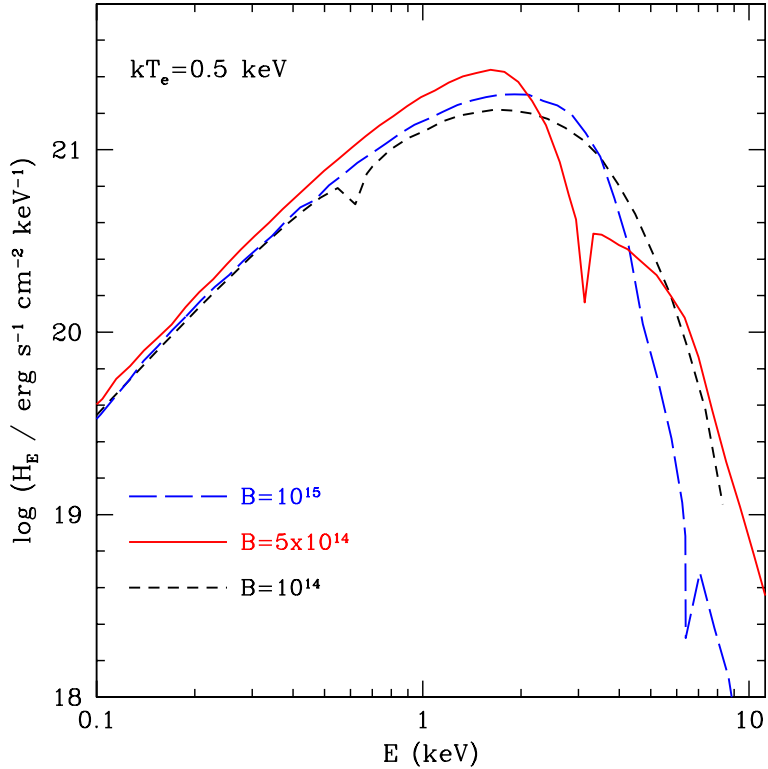,width=11truecm} } \figcaption[]{The
spectra of radiation emerging from a strongly magnetized neutron star
atmosphere of $T_e=0.5$~keV, when proton cyclotron and vacuum
resonances have been taken into account. \label{Fig:spec}}
\end{figure}

We now discuss the effect of the vacuum polarization and proton
cyclotron resonances on the spectra of emission from the surface of a
strongly magnetized neutron star. Computing the spectrum of surface
radiation requires the solution of the two coupled equations that
describe the propagation of the photons through the magnetized plasma,
subject to the condition of radiative equilibrium in a hydrostatic
atmosphere. In the calculations presented here, we follow the methods
described in \"Ozel (2001) with the addition of the new algorithm for
handling resonances discussed in the previous section. Briefly, we use
a modified Feautrier method for the solution of the angle- and
polarization-mode dependent radiative transfer problem and ensure
radiative equilibrium with a temperature correction scheme based on
the Lucy-Uns\"old algorithm. The implementations of these methods are
given in \"Ozel (2001).

Figure~\ref{Fig:spec} shows the spectra of surface emission from a
neutron star with $B=10^{14}-10^{15}$~G and with an effective
temperature of $\te=0.5$~keV. The spectra at all field strengths are
harder than a blackbody at $\te =0.5$~keV. However, the shape of the
continuum as well as the narrow features that appear due to the proton
cyclotron resonance depend strongly on the magnetic field. 

The shape of the continuum is determined by both the vacuum
polarization and the proton cyclotron resonances. As discussed in
\"Ozel (2001), the vacuum resonance introduces a layer of enhanced
interactions and brings the thermalization depth of all photons with
$E \gtrsim 2$~keV closer to the outermost layers of the atmosphere.
The sudden increase in the opacity at these photon-energy dependent
critical densities leads to broad-band absorption-like features in the
spectrum. However, because this resonance has a strong dependence on
photon energy that causes less attenuation of the flux at higher
photon energies, it leads to a hardening of the spectrum at $E \gtrsim
2$~keV for $B \gtrsim 10^{14}$~G as well as to increased flux at
photon energies $E \lesssim 1$~keV, as discussed in \"Ozel (2001) and
Bulik \& Miller (1997). Note that the resonance does not soften the
spectrum but indeed is {\it responsible} for the hard tails at $E >
2$~keV.

The spectra in Figure~\ref{Fig:spec}, however, do not all show these
hard tails because the proton cyclotron resonance also affects the
shape of the continuum through its modification of the ellipticities
of the normal modes and its contribution to the opacities. In
particular, it gives rise to a broad absorption feature that reduces
the continuum flux around the proton cyclotron energy for $B \gtrsim
{\rm few} \times 10^{14}$~G.  For $B \sim 5\times 10^{14}$~G, the
cyclotron absorption modifies the peak of the spectrum, leading to a
sharper fall-off at $E \sim 2$~keV, while for $B \sim 10^{15}$~G, it
suppresses the spectral tail in the $5-10$~keV energy range.

The narrow features at the proton cyclotron energy, on the other hand,
arise from the enhanced interaction of the protons with the photons at
that energy. These features are weak in the presence of the vacuum
polarization resonance: the thermalization depth of the photons in
nearby energies are brought closer to the thermalization depth at the
cyclotron energy because of the vacuum polarization resonance, thus
reducing the contrast between the flux at these photon energies.  The
resulting features have small equivalent widths at all field strengths
considered here: at $B=5\times10^{14}$~G, the equivalent width is
$\sim 80$~eV, while at $B=10^{15}$~G, the equivalent width is $\sim
0.2$~keV. These features are likely to have even smaller equivalent
widths as seen by an observer at infinity because of the effects of
phase averaging and redshifts on the observed spectra.

\section{Conclusions}

In this paper, we have considered the various effects of vacuum
polarization and proton cyclotron resonances on the propagation of
photons through a strongly magnetized plasma. Given that the treatment
of photon transport by solving two coupled transfer equations for two
normal modes of propagation assumes large Faraday depolarization, we
have first checked whether this condition is satisfied at the resonant
densities for all photon energies. We have found that for photon
energies $E \gtrsim 1$~keV, the assumption is satisfied, i.e., the
modes evolve adiabatically through the resonance, for all directions
of propagation at the magnetic field strengths considered here.  The
resonant layer for photons with smaller energies lies deeper in the
atmosphere than their thermalization depths and thus does not affect
their propagation. Employing a normal mode treatment to calculate the
properties of radiation emerging from a magnetized plasma is therefore
justified.

We have then constructed radiative equilibrium atmosphere models of
strongly magnetized neutron stars that includes the effects of vacuum
polarization and proton cyclotron resonances. We have introduced a new
numerical method that resolves accurately the sharp changes of the
absorption and mode-coupling cross sections at the resonant
densities. In particular, this method involves sampling the resonance
region with a very large number of points on an auxiliary grid and
thus allows for an accurate computation of the total optical depth
across the resonances. Using this method in addition to those
described in \"Ozel (2001) for the solution of the transfer problem
subject to the radiative equilibrium condition, we have calculated the
spectral energy distributions of a cooling neutron star atmosphere.

We have shown that the resulting spectra are harder at all magnetic
field strengths than a blackbody at the effective temperature but the
shape of the continuum depends strongly on the field strength, shaped
by the broad-band absorption due to both resonances. In particular,
the suppression of the flux due to the proton cyclotron resonance
dramatically reduces the hard tails at $E > 3$~keV that arise from the
vacuum resonance at $B = 10^{15}$~G. In contrast, it only modifies the
peak of the spectrum at $B = 5 \times 10^{14}$~G and, together with
the vacuum resonance, gives rise to a hard tail of photon index
$\Gamma \approx 2-3$ in the $2-6$~keV range. Note that hard tails have
been observed in this energy range in a number of sources that are
currently the best candidates for being ultramagnetized neutron stars,
such as the anomalous X-ray pulsars and soft gamma-ray
repeaters. However, a conclusive statement can be made by comparing
the observations of these sources to spectral models that take into
account the effects of the gravitational redshift and phase-averaging.

Finally, we have shown that the narrow absorption features introduced
by the proton cyclotron resonance have small equivalent widths. The
vacuum polarization resonance significantly suppresses the strengths
of these lines by bringing the thermalization depth of photons that
have energies in the vicinity of the cyclotron energy further out in
the atmosphere and thus closer to that of the photons at the cyclotron
energy. These small equivalent widths may help explain the lack of
narrow absorption features in the observations of anomalous X-ray
pulsars and soft gamma-ray repeaters (Juett et al.\ 2001; Patel et
al.\ 2001). Deeper observations of these sources, especially with
instruments that have higher sensitivity at $E \gtrsim 5$~keV range
may help reveal some of these features and thus the nature of these
intriguing objects.

\acknowledgements

I thank John Bahcall for his valuable input and discussions on the
physics of mode evolution across resonances. I also thank Dimitrios
Psaltis for many discussions on the treatment of resonances in
radiative transfer problems and Ramesh Narayan for useful
suggestions. This work was supported in part by a fellowship of the
Keck Foundation and an NSF grant PHY-0070928.

\appendix

\section{On the Adiabatic Mode Evolution and Mode Conversion
near the Vacuum Polarization Resonance}

In a recent paper, Lai \& Ho (2002) discussed the physics of vacuum
polarization resonance and pointed out the well-known effects of
adiabatic evolution and enhanced polarization mode conversion that
take place through the resonant density (see the references in \S
1). They argued that this effect was not treated in previous studies
and evaluated the conditions under which the large Faraday
depolarization assumption and hence the adiabatic evolution of modes
breaks down. In a subsequent work, Ho \& Lai (2002) further claimed
that they included for the first time the effect of this new
phenomenon on the spectra of a magnetized neutron star atmosphere in
radiative equilibrium.

In this appendix, we clarify the effects of vacuum polarization
resonance on the normal-mode description of photon transport in
magnetized media. We show that the effects discussed by Lai \& Ho
(2002) have been taken into account in the previous calculations of
photon transport through a plasma (Bulik \& Miller 1997; Zane et al.\
2000; \"Ozel 2001). In particular, as long as a normal-mode treatment
is employed, the physics included in the calculations is the same,
independent of the nomenclature with which one describes modes above
and below the resonant density. Below, we outline some of the
mistakes and inconsistencies in their discussion.

First, referring to the case where the photon modes evolve
adiabatically as ``mode conversion'' is misleading, since this is
precisely the case where the normal modes of propagation ($i=1,2$ or
$-/+$ modes) remains the same above and below the resonance. What {\it
is\/} different is the correspondence between the polarization and
propagation eigenstates on the two sides of the resonance.  Note in
particular that equations (2.27) and (2.43) of Ho \& Lai (2002), which
presumably describe the two different definitions of normal modes, are
mathematically identical.

Second, adiabatic evolution is not an additional effect that needs to
be {\it included\/} in calculations of radiative transport; it is by
definition part of the calculations.  In fact, as we discussed in \S
2, solving the transfer equation in the two normal modes {\it
requires\/} that the modes evolve adiabatically. What the modes are
called above and below the resonant density is irrelevant, given that
the transfer equations are written for the normal modes of propagation
and not for polarization eigenstates (the so-called extraordinary and
ordinary modes) as we discussed above. Therefore, they criticise
incorrectly the previous works for not including this new physical
effect. For the same reasons, it is in fact meaningless to ``include''
or to ``neglect'' mode conversion (in the terminology of Lai \& Ho
2002) because neglecting the adiabatic mode evolution does not
describe any physical situation. The difference in the results between
these two cases most likely arises from the different numerical
treatments the authors employ in each case, both of which are highly
inaccurate as discussed in \S 4.

Ho \& Lai (2002, \S 2.4) also claim that in the limit of
non-adibaticity, the radiative transfer formalism breaks down and
cannot describe the evolution of photons. This is also incorrect. As
discussed above, one simply needs to solve all four equations
(\ref{eq:full}) in this case rather than the two equations for the
diagonal terms of $R_{ij}$. We emphasize that this case is not
relevant for the problem discussed here but in general can be easily
addressed by keeping the equations for the off-diagonal terms.


\begin{references}

\reference{adler} Adler, S.\,L.\ 1971, Ann.\ Phys., 67, 599

\reference{pavlov} Bezchastnov, V.~G., Pavlov, G.~G., Shibanov,
Y.~A., \& Zavlin, V.~E.\ 1996, Gamma-Ray Bursts, Proceedings of the
3rd Huntsville Symposium (Woodbury, AIP) eds.  C. Kouveliotou, M.\
F.~Briggs, and G.\ J.~Fishman, 907

\reference{bulik} Bulik, T.~\& Miller, M.~C.\ 1997, \mnras, 288, 596

\reference{gp74} Gnedin, I.~N.~\& Pavlov, G.~G.\ 1974, Zhurnal
Eksperimental noi i Teoreticheskoi Fiziki, 65, 1806

\reference{hl01} Ho, W.~C.~G.~\& Lai, D.\ 2001, \mnras, 327, 1081

\reference{HL02} Ho, W.~C.~G.~\& Lai, D.\ 2002, \mnras, submitted
(astro-ph/0201380)

\reference{juett} Juett, A.~M., Marshall, H.~L., Chakrabarty, D., \&
Schulz, N.~S.\ 2002, \apjl, 568, L31

\reference{kps} Kaminker, A.~D., Pavlov, G.~G., \& Shibanov, I.~A.\
1982, \apss, 86, 249

\reference{LH02} Lai, D.~\& Ho, W.~C.~G.\ 2002, \apj, 566, 373

\reference{meszaros} M\'esz\'aros, P.\ 1992, High-Energy Radiation from
Magnetized Neutron Stars (Chicago: University Press)

\reference{mv79} Meszaros, P.~\& Ventura, J.\ 1979, \prd, 19, 3565

\reference{ozel01} {\" O}zel, F.\ 2001, \apj, 563, 276

\reference{patel} Patel, S.~K.~et al.\ 2001, \apjl, 563, L45

\reference{1978SvA....22..214P} Pavlov, G.\ G.\ \& Shibanov, I.\ A.\ 
1978, Sov.\ Astr., 22, 214

\reference{PS79} ---------.\ 1979, Sov.\ Phys.\ JETP, 49, 741

\reference{te} Tsai, W.~\& Erber, T.\ 1975, \prd, 12, 1132

\reference{Zane} Zane, S., Turolla, R., Stella, L., \& Treves, A.\ 2001, 
\apj, 560

\end{references}
\end{document}